\documentclass[12pt]{article}

\usepackage{amsmath}
\usepackage{amsfonts}
\usepackage{amssymb}

\newcommand{\be}{\begin{equation}}
\newcommand{\ee}{\end{equation}}
\newcommand{\ba}{\begin{eqnarray}}
\newcommand{\ea}{\end{eqnarray}}

\begin{document}
\begin{flushright}
{\tt Vestnik Leningradskogo Universiteta (Ser.4 Fiz.Khim.) 4(25) (1988) pp.3-9 (translated from Russian)}
\end{flushright}
\vspace{1.0cm}

\begin{center}
{\large\bf FACTORIZATION METHOD IN CURVILINEAR COORDINATES AND PAIRING OF LEVELS FOR MATRIX POTENTIALS}\\

\vspace{0.3cm} {\large \bf A.A.Andrianov\footnote{e-mail: sashaandrianov@gmail.com}, M.V. Iof\/fe\footnote{e-mail: m.ioffe@pobox.spbu.ru},
Tsu Zhun-Pin
}\\
\vspace{0.2cm}
Saint-Petersburg (Leningrad) State University,
198504 St.-Petersburg, Russia\\
\end{center}
\hspace*{0.5in}
\begin{minipage}{5.0in}
{\small Multidimensional factorization method is formulated in arbitrary curvilinear coordinates. Particular cases
of polar and spherical coordinates are considered and matrix potentials with separating variables are constructed.
A new class of matrix potentials is obtained which reveals a double degeneracy or equidistant splitting of energy levels
(hidden symmetry).
\\
\vspace*{0.1cm}
}
\end{minipage}

\vspace*{0.4cm}

\setcounter{footnote}{0} \setcounter{equation}{0}
\section{Introduction.}

The Factorization Method is an effective tool in search and construction of exactly solvable Schr\"odinger problems \cite{1} and of the
models with equivalent (almost coinciding) energy spectra \cite{2}, \cite{3}. The mainstream of the method was related with one-dimensional problems \cite{1}.

Recently, the multidimensional generalization of the Factorization Method was elaborated \cite{4} - \cite{6}, and its supersymmetric origin was demonstrated \cite{7}, \cite{8}. The main results for multidimensional case were obtained in Cartesian coordinates. Meanwhile, the symmetry properties of original potential mark out the preferred coordinate systems related to orbits of the symmetry group (separation of variables). Usually, these coordinate systems are curvilinear ones.

Thus, it would be interesting to reformulate the Multidimensional Factorization Method in curvilinear coordinates and to diagonalize matrix potentials appeared under the multidimensional Darboux transformation (Sect. 2). Thereby, starting from scalar potentials with separated variables a class of matrix potentials will be found which is also amenable to separation of variables . The polar and spherical coordinates and the particular case of Coulomb potential will be considered below in detail.

The full variety of Hamiltonians interrelated by the intertwining relations obeys twofold degeneracy of the spectrum. The latter has \cite{2}, \cite{3}, \cite{7}, \cite{8} the supersymmetric interpretation: spectra of systems with even and odd fermionic modes coincide.

However, the Factorization Method allows the construction of potentials with additional twofold level degeneracy without a simple supersymmetric interpretation. Together with standard supersymmetry of the full system of equivalent Hamiltonians, this means fourfold degeneracy of levels. Such potentials are built in Sect. 3, and the Factorization Method in curvilinear coordinates is more convenient for their description. More general class of potentials whose energy levels are paired with a given constant (equidistant) splitting is found.

\section{Factorization Method in curvilinear coordinates.}

We shall start from the case of two-dimensional space with arbitrary curvilinear coordinates $(q^1, q^2),$ where
the interval $ds^2=g_{ik}dq^idq^k$ is given by the corresponding metric tensor $g_{ik}.$

The required generalization of Factorization Method includes:

1) constructing the operators $Q_l^{\pm}\,(l=1,2),$ which factorize the initial Hamiltonian:
$$H^{(0)}=-\frac{1}{2}\triangle + V^{(0)};$$

2) constructiing the matrix Hamiltonian
\begin{equation}\label{1}
H_{ik}^{(1)}=-\frac{1}{2}\triangle + V^{(1)}_{ik},
\end{equation}
which is connected with $H^{(0)}$ by intertwining relations;

3) constructing the scalar Hamiltonian
$$H^{(0)}=-\frac{1}{2}\triangle + V^{(0)},$$
also intertwined with $H^{(1)}_{ik}.$

These necessary steps of the Factorization Method provide the well known links \cite{4} - \cite{6} between the
spectra of operators $H^{(0)},\, H^{(1)}_{ik},\, H^{(2)}$ and between the corresponding eigenfunctions
$\Psi^{(0)},\, \Psi^{(1)},\,\Psi^{(2)}.$

The expression for the operators $Q^{\pm}$ can be found by using the representation of the Laplace operator (kinetic terms in the Hamiltonians above)
in arbitrary curvilinear coordinates:
$$\triangle = g^{ik}\nabla_i\nabla_k,$$
where $g^{ik}$ is a contravariant tensor inverse to the metric tensor $g_{ik},$ and the covariant derivative operator $\nabla_i$
is defined by means of three-index Christoffel symbols $\Gamma :$
\ba
\nabla_i\varphi &\equiv &\partial_i\varphi\equiv \frac{\partial}{\partial q^i}\varphi \quad\,\quad\quad\quad\quad\quad (\varphi - scalar),\nonumber\\
\nabla_i A_j &\equiv &\partial_i A_j-\Gamma^k_{ji}A_k \quad\quad\quad\quad\quad\,\, (A_j - covariant\,\, vector), \nonumber\\
\nabla_i A^j &\equiv &\partial_i A^j+\Gamma^j_{ki}A^k \quad\quad\quad\quad\quad\,\, (A^j - contravariant\,\, vector), \label{2}\\
\nabla_i B_{jl} &\equiv &\partial_i B_{jl}-\Gamma^k_{ji}B_{kl} - \Gamma^k_{li}B_{jk}  \,\quad (B_{jl} - second\,\, rank\,\, covariant\,\, tensor),\nonumber
\ea
etc. The Christoffel symbols are expressed in terms of the metric tensor $\Gamma^k_{ij}=\frac{1}{2}g^{kl}\biggl(\frac{\partial g_{il}}{\partial q^j}
+\frac{\partial g_{jl}}{\partial q^i}-\frac{\partial g_{ij}}{\partial q^l}\biggr),$ and they are introduced in differential calculus
of tensor objects to take into account the change of basic vectors and metrics under parallel transport .
In a particular case of Cartesian orthogonal coordinates $g_{ij}=\delta_{ij},\,\, \Gamma^k_{ij}=0,\,\, \nabla_i=\partial_i.$

Let us choose the operators $Q^{\pm}_l$ in the form:
\be
Q^{\pm}_l\equiv \frac{1}{\sqrt{2}}\biggl(\mp\nabla_l+(\nabla_l\chi )\biggr)= \frac{1}{\sqrt{2}}\biggl(\mp\nabla_l+(\partial_l\chi)\biggr),
\label{3}
\ee
where $\exp{(-\chi )}-$ one of the two solutions of the Schr\"odinger equation $H^{(0)}\exp{(-\chi )}=\overline{E}\exp{(-\chi )},$ and
$$Q^{l\,\pm}=g^{lk}Q_k^{\pm},\quad [Q_l^-,\, Q_k^+]=(\nabla_l\nabla_k\chi ),\quad [Q_l^-,\,Q_k^-]=[Q^+_l,\,Q_k^+]=0.$$
The presence of the volume factor $\sqrt{\det(g_{ik})}$ in the scalar product and
the property $\nabla_ig_{kj}=0$ are necessary to prove that operators $Q^{\pm}_l$ are mutually conjugate\footnote{We remind that the flat space only is considered in this paper.}.

Then, $H^{(0)}$ can be written in a factorized form:
\be
H^{(0)}=Q_l^+Q^{l\,-}+\overline{E}=-\frac{1}{2}\triangle + \frac{1}{2}\biggl[(\nabla_l\chi)(\nabla^l\chi)-(\nabla_l\nabla^l\chi)\biggr]+\overline{E},
\label{4}
\ee
where the covariant derivative in $Q_l^-$ acts onto a scalar $\nabla_l\Psi=\partial_l\Psi ,$ and covariant derivative in $Q_l^+$ acts
onto a vector in accordance with (\ref{2}).

To build the matrix Hamiltonian $H^{(1)}$ let's use the operators $P_l^{\pm}\equiv\sqrt{g}\epsilon_{lk}Q^{k\, \mp}, \quad P^{l\, \pm}=g^{lk}P^{\pm}_k=
\frac{1}{\sqrt{g}}\epsilon^{lk}Q_k^{\mp}\,\,(g\equiv\det{g_{ik}},$ with $\epsilon_{lk}$ being completely antisymmetric unit pseudotensor).
The important property of orthogonality $P_k^+Q^{k\, -}=P_k^-Q^{k\, +}=Q_k^+P^{k\, -}=Q_k^-P^{k\, +}=0$ guarantees, in particular, intertwining the operator $H^{(0)}$ and  the operator $H^{(1)\,k}_{i}:$
\ba
&&H^{(1)\,k}_{l}\equiv Q_l^-Q^{k\, +}+P_l^-P^{k\, +}+ \delta_l^k\overline{E}=\delta_l^kH^{(0)}+(\nabla_l\nabla^k\chi ) \label{5}\\
&&H^{(1)\, k}_l Q_k^-=Q_l^- H^{(0)};\quad Q^{l\,+}H^{(1)\, k}_l=H^{(0)}Q^{k\, +}. \nonumber
\ea
Similarly to the case of Cartesian coordinates \cite{4} - \cite{6}, these relations lead to the connection of spectra of $H^{(0)}$
and $H^{(1)}:\,\,\{E_n^{(0)}\} \subset \{E_n^{(1)}\}$ and of their wave functions:
$$\Psi_k^{(1)}(E)=\frac{4}{\sqrt{E-\overline{E}}}Q_k^-\Psi^{(0)}(E),\,\, \Psi^{(0)}(E)=\frac{1}{\sqrt{E-\overline{E}}}Q^{k\, +}\Psi^{(1)}_k(E).$$
Analogously, the operator $H^{(1)\, k}_l$ is intertwined with the scalar Hamiltonian:
\be
H^{(2)}\equiv P_l^+P^{l\, -}+\overline{E}=-\frac{1}{2}\triangle + \frac{1}{2}\biggl[(\nabla_l\chi)(\nabla^l\chi)+(\nabla_l\nabla^l\chi )\biggr]
+\overline{E}=H^{(0)}+\triangle\chi,
\label{6}
\ee
whose spectrum $\{E_n^{(2)}\}$ also lies in the spectrum of $H^{(1)\, k}_l.$ An arbitrary point of the spectrum of $H^{(1)}$ coincides
either with $E_n^{(0)}$ or with $E_n^{(2)}:$
$$H^{(1)\, k}_lP_k^-=P_l^-H^{(2)};\quad P^{l\, +}H^{(1)\, k}_l=H^{(2)}P^{k\, +}$$
$$\Psi_k^{(1)}(E)=\frac{1}{\sqrt{E-\overline{E}}}P_k^-\Psi^{(2)}(E); \quad \Psi^{(2)}(E)=
\frac{1}{\sqrt{E-\overline{E}}}P^{k\, +}\Psi_k^{(1)}(E).$$

For illustration we consider the case of polar coordinates $(\rho,\,\varphi)$ on the plane:
$$
g_{ij}=
\left(
  \begin{array}{cc}
    1 & 0 \\
    0 & \rho^2 \\
  \end{array}
\right);\quad \Gamma^1_{22}=-\rho;\quad \Gamma^2_{12}=\Gamma^2_{21}=\frac{1}{\rho}.
$$
From (\ref{4}) - (\ref{6}) one obtains the explicit form of potentials $(\partial_1\equiv \partial / \partial\rho,\,
\partial_2\equiv\partial / \partial\varphi):$
\ba
V^{(0)}&=&\frac{1}{2}\biggl[(\partial_1\chi)^2+\frac{1}{\rho^2}(\partial_2\chi)^2-(\partial_1^2\chi)-
\frac{1}{\rho^2}(\partial_2^2\chi)-\frac{1}{\rho}(\partial_1\chi)\biggr]+\overline{E},
\label{7}\\
V^{(1)\, k}_i&=&\delta^k_iV^{(0)}+(\nabla_i\nabla^k\chi )=
\left(
\begin{array}{cc}
V^{(0)}+\partial_1^2\chi & \frac{1}{\rho^2}\partial_1\partial_2\chi-\frac{1}{\rho^3}\partial_2\chi \\
\partial_1\partial_2\chi-\frac{1}{\rho}\partial_2\chi & V^{(0)}+\frac{1}{\rho^2}\partial_2^2\chi +\frac{1}{\rho}\partial_1\chi\\
\end{array}
\right),
\label{8}\\
V^{(2)}&=&\frac{1}{2}\biggl[(\partial_1\chi)^2+\frac{1}{\rho^2}(\partial_2\chi)^2+(\partial_1^2\chi)+
\frac{1}{\rho^2}(\partial_2^2\chi)+\frac{1}{\rho}(\partial_1\chi)\biggr]+\overline{E}.
\label{9}
\ea
For the particular case of centrally symmetrical $\chi =\chi (\rho),$ the matrix potential is diagonal just
in polar coordinates:
$$
V^{(1)\, k}_i=
\left(
  \begin{array}{cc}
    V^{(0)}(\rho)+\partial_1^2\chi + \overline{E} & 0 \\
    0 & V^{(0)}(\rho)+\frac{1}{\rho}\partial_1\chi + \overline{E} \\
  \end{array}
\right).
$$

The method can be generalized to a space of arbitrary dimension. For the case of physically
interesting three-dimensional space, the operators $P^{\pm}$ are second rank tensors:
$$
P_{ik}^{\pm}=\sqrt{g}\epsilon_{ikl}Q^{l\, \mp},\quad P_{ik}^+Q^{k\, -}=Q^{k\, +}P_{ik}^-= P_{ik}^-Q^{k\, +}
=Q^{k\, -}P_{ik}^+=0.
$$
The intertwining relations are:
\ba
&&H^{(1)\, k}_i Q_k^-=Q_i^-H^{(0)};\quad Q_k^+H^{(1)\, k}_i=H^{(0)}Q_i^+;\quad H_i^{(2)\, k}P_{kl}^+=P_{im}^+H^{(1)\, m}_l;\nonumber\\
&&P_{lk}^-H^{(2)\, k}_i=H^{(1)\, m}_lP^-_{im};\quad H^{(3)}Q^{k\, -}=Q^{i\, -}H^{(2)\, k}_i;\quad Q_i^+H^{(3)}=H^{(2)\, k}_i Q_k^+,\nonumber
\ea
where
\ba
H^{(0)}&=&Q_l^+Q^{l\, -}+\overline{E}=-\frac{1}{2}\triangle+\frac{1}{2}\biggl[(\nabla_l\chi)(\nabla^l\chi)-\triangle\chi \biggr];\nonumber\\
H^{(1)\, k}_i&=&Q_i^-Q^{k\, +}+P_{li}^-P^{lk\, +}+\overline{E}=\delta_i^kH^{(0)}+(\nabla_i\nabla^k\chi );\nonumber\\
H^{(2)\, k}_i&=&Q_i^+Q^{k\, -}+P_{li}^+P^{kl\, -}+\overline{E}=\delta_i^kH^{(3)}-(\nabla_i\nabla^k\chi );\nonumber\\
H^{(3)}&=&Q_l^-Q^{l\, +}+\overline{E}=-\frac{1}{2}\triangle+\frac{1}{2}\biggl[(\nabla_l\chi)(\nabla^l\chi)+\triangle\chi \biggr].\nonumber
\ea
For the particular case of spherical coordinates $(r,\, \theta ,\, \varphi)$, one has:
$$
g_{11}=1;\quad g_{22}=r^2\quad g_{33}=r^2\sin^2\theta ,
$$
$$
\Gamma_{22}^1=-r;\quad \Gamma_{33}^1=-r\sin^2\theta;\quad \Gamma_{33}^2=-\sin\theta\cos\theta;\quad \Gamma_{12}^2=\Gamma_{13}^3=\frac{1}{r};
\quad \Gamma_{23}^3=\cot\theta,
$$
and
\ba
V^{(0)}&&=\frac{1}{2}\biggl[(\partial_1\chi)^2+\frac{1}{r^2}(\partial_2\chi)^2+\frac{1}{r^2\sin^2\theta}(\partial_3\chi)^2-\partial^2_1\chi-
\frac{2}{r}\partial_1\chi-\frac{1}{r^2}\partial_2^2\chi-\frac{\cot\theta}{r^2}\partial_2\chi-\nonumber\\
&&-\frac{1}{r^2\sin^2\theta}\partial_3^2\chi\biggr]
+\overline{E},\nonumber\\
V^{(1)\, k}_i&&=\delta^k_iV^{(0)}+\nabla_i\nabla^k\chi=\nonumber\\
&&=\left(
  \begin{array}{ccc}
    V^{(0)}+\partial_1^2\chi & \frac{1}{r^2}\biggl(\partial_1\partial_2\chi-\frac{1}{r}\partial_2\chi\biggr) &\frac{1}{r^2\sin^2\theta}\biggl(\partial_1\partial_3\chi-\frac{1}{r}\partial_3\chi\biggr)  \\
    \partial_1\partial_2\chi-\frac{1}{r}\partial_2\chi & V^{(0)}+\frac{1}{r^2}\partial_2^2\chi+\frac{1}{r}\partial_1\chi &\frac{1}{r^2\sin^2\theta}\biggl(\partial_2\partial_3\chi-\cot\theta\partial_3\chi\biggr)  \\
    \partial_1\partial_3\chi-\frac{1}{r}\partial_3\chi & \frac{1}{r^2}\biggl(\partial_2\partial_3\chi-\cot\theta\partial_3\chi\biggr) & V^{(0)}+\frac{1}{r^2\sin^2\theta}\partial_3^2\theta+\frac{1}{r}\partial_1\chi+\frac{\cot\theta}{r^2}\partial_2\chi \\
  \end{array}
\right),\nonumber\\
V^{(2)\, k}_i&&=V^{(1)\, k}_i\left(
                                      \begin{array}{c}
                                        \chi\rightarrow -\chi \\
                                        V^{(0)}(\chi )\rightarrow V^{(3)}(\chi ) \\
                                      \end{array}
                                    \right),\nonumber\\
V^{(3)}&&=V^{(0)}(\chi \rightarrow -\chi).\nonumber
\ea
From these expressions, one can conclude that in terms of the spherical coordinates in 3-dimensional
space the matrix potentials $V^{(1)},\, V^{(2)}$ are also diagonal for spherically symmetrical case $\chi = \chi(r).$ Thus, the scalar problem amenable to separation of variables produces the matrix problems which allow the separation of variables as well.

From this point of view, the problem of the Coulomb potential and its Darboux transformation, considered earlier in \cite{9}, looks interesting:
\ba
V^{(0)}&=&-\frac{\alpha}{r};\quad V^{(3)}=+\frac{\alpha}{r}; \nonumber\\
V^{(1)\, k}_i&=&\left(
                  \begin{array}{ccc}
                    -\frac{\alpha}{r} & 0 & 0 \\
                    0 & 0 & 0 \\
                    0 & 0 & 0 \\
                  \end{array}
                \right);\quad
                V^{(2)\, k}_i = \left(
                  \begin{array}{ccc}
                    +\frac{\alpha}{r} & 0 & 0 \\
                    0 & 0 & 0 \\
                    0 & 0 & 0 \\
                  \end{array}
                \right). \label{10}
\ea

Thus, the Factorization Method developed above in curvilinear coordinates allows to
unravel the structure of the matrix Coulomb potential. When interpreting $\Psi^{(1)}_i$ as a vector particle wave function, then it follows from
(\ref{10}) that bound states in matrix Coulomb potential exist for one polarization (along $r$) only, and the motion with
two other polarizations is free.

\section{Matrix potentials with pairing of levels.}

The interrelation of spectra of the Hamiltonians $H^{(n)}$ for arbitrary space dimension $d$ was studied in \cite{4} - \cite{6}, and it was
proven
there that the spectrum of the joint Hamiltonian
\ba
\widehat H =
\left(
  \begin{array}{cccc}
    H^{(0)} & 0 & 0 & 0 \\
    0 & H^{(1)} &  & 0 \\
    0 & 0 & ... & 0 \\
    0 & 0 & 0 & H^{(d)} \\
  \end{array}
\right)
\label{11}
\ea
is twofold degenerate (may be, excluding the ground state). This degeneracy has \cite{7}, \cite{8} a supersymmetric interpretation:
the spectra of Hamiltonians coincide for even and odd number of fermionic excitations:
\ba
H_B=\left(
      \begin{array}{ccc}
        H^{(0)} & 0 & 0 \\
        0 & H^{(2)} & 0 \\
        0 & 0 & ... \\
      \end{array}
    \right);\quad
    H_F=\left(
      \begin{array}{ccc}
        H^{(1)} & 0 & 0 \\
        0 & H^{(3)} & 0 \\
        0 & 0 & ... \\
      \end{array}
    \right).\nonumber
\ea
At the same time, a class of potentials with higher degree of degeneracy exists. In order to construct the potentials with additional degeneracy in 2-dimensional space, we examine the case with identical potentials $V^{(0)}$ and $V^{(2)}:\quad \delta V\equiv V^{(2)}-V^{(0)}=\triangle\chi =0.$
The real solutions of the Laplace equation are simply expressed in Cartesian coordinates in terms of arbitrary analytical functions of complex variable $z\equiv x+iy: \quad \chi = F(z)+\overline{F(z)}=2Re F(z),\,\, \bar{z}\equiv x-iy.$

However, polar coordinates are more convenient to study the physical properties of potentials $V^{(n)}.$ The general solution of the Laplace equation, nonsingular for finite $\rho ,$ is:
\be
\chi = \sum_{m=0}^{\infty}\alpha_m\chi_m(\rho, \varphi),\quad \chi_m=\rho^m\sin (m\varphi + \delta_m),
\label{12}
\ee
where $\alpha_m,\,\delta_m $ are arbitrary real numbers. Substituting these solutions into (\ref{7}) -  (\ref{9}), one obtains a class of nonsingular potentials:
\be
V^{(0)}=V^{(2)}=\frac{1}{2}\sum_{m=1}^{\infty}m^2\alpha^2\rho^{2m-2}+\sum_{n<m=1}^{\infty}mn\alpha_m\alpha_n\rho^{m+n-2}\cos [(m-n)\varphi+\delta_m-\delta_n],
\label{13}
\ee
\be
V^{(1)\, k}_i=\left(
                \begin{array}{cc}
                  V^{(1)\, 1}_1 & \sum_{m=2}^{\infty }m(m-1)\alpha_m\rho^{m-3}\cos (m\varphi +\delta_m) \\
                  \sum_{m=2}^{\infty }m(m-1)\alpha_m\rho^{m-1}\cos (m\varphi +\delta_m)  & V^{(1)\, 2}_2 \\
                \end{array}
              \right).
\label{14}
\ee
where
$$V^{(1)\, 1}_1=V^{(0)}+\sum_{m=2}^{\infty}m(m-1)\alpha_m\rho^{m-2}\sin (m\varphi +\delta_m);$$
$$V^{(1)\, 2}_2=V^{(0)}-\sum_{m=2}^{\infty}m(m-1)\alpha_m\rho^{m-2}\sin (m\varphi +\delta_m).$$
The asymptotic behavior of (\ref{13}) and (\ref{14}) for large $\rho$ (and arbitrary $\varphi $) is determined by
the first term in (\ref{13}), and it guarantees the existence of bound states. The particular cases of potential
$V^{(0)}$ of this kind are the harmonic potential $V^{(0)}=\alpha\rho^2$ and the  anharmonic potential of the form
$V^{(0)}=\alpha\rho^2+\beta\rho^4+4\alpha\beta\rho^2\cos(\varphi +\delta ).$

Due to coincidence of potentials $V^{(0)}$ and $V^{(2)},$ the wave functions of matrix Hamiltonian $H^{(1)\, k}_i,$
built by means of 2-dimensional Darboux transformation from the wave functions of scalar Hamiltonians $H^{(0)}$
and $H^{(2)}$, read,
$$
\Psi_i^{(1)}=\frac{1}{\sqrt{E_n^{(0)}-\overline{E}}}Q_i^-\Psi^{(0)}(E_n^{(0)}),\quad
\widetilde\Psi_i^{(1)}=\frac{1}{\sqrt{E_n^{(0)}-\overline{E}}}P_i^-\Psi^{(2)}(E_n^{(2)}\equiv E_n^{(0)}),
$$
and they have the same eigenvalue $E_n^{(1)}=E_n^{(0)}=E_n^{(2)}.$ In other words all energy levels in potential $V^{(1)\, k}_i$
are twofold degenerate. Coincidence of these levels with levels of $H^{(0)}$ and $H^{(2)}$ means that the joint Hamiltonian
$$
\widehat H=\left(
             \begin{array}{ccc}
               H^{(0)} & 0 & 0 \\
               0 & H^{(1)} & 0 \\
               0 & 0 & H^{(2)} \\
             \end{array}
           \right)
$$
has fourfold degeneracy of all levels.

A more general problem corresponds to $\delta V=V^{(2)}-V^{(0)}=\varepsilon = Const .$ In such a case, evidently $E_n^{(2)}=E_n^{(0)}+\varepsilon ,$
leading to related equidistant splitting of all levels of the matrix Hamiltonian $H^{(1)\, k}_i :$
$$
E_{2n-1}^{(1)}=E_{n}^{(0)},\,\,\, E_{2n}^{(1)}=E_{n}^{(2)}=E_{n}^{(0)}+\varepsilon .
$$
The explicit form of such potentials is obtained by solving the equation $\triangle\chi=\varepsilon :$
$$
\chi = \frac{1}{4}\varepsilon\rho^2 + \sum_{m=0}^{\infty}\alpha_m\chi_m(\rho, \varphi),
$$
where $\chi_m$ is given by (\ref{12}). Then,
\ba
V^{(2)}&=&V^{(0)}+\varepsilon=\frac{1}{8}\varepsilon^2\rho^2+
\frac{1}{2}\sum_{m=1}^{\infty}m^2\alpha_m^2\rho^{2m-2}+\nonumber\\
&+&
\sum_{n<m=1}^{\infty}mn\alpha_m\alpha_n\rho^{m+n-2}\cos [(m-n)\varphi +\delta_m-\delta_n]+\frac{1}{2}\varepsilon\sum_{m=1}^{\infty}m\alpha_m\rho^m\sin(m\varphi +\delta_m)+\frac{1}{2}\varepsilon,
\nonumber
\ea
$$
V^{(1)\,k}_i=\left(
               \begin{array}{cc}
                  V^{(1)\, 1}_1 & \sum_{m=2}^{\infty}m(m-1)\alpha_m\rho^{m-3}\cos(m\varphi +\delta_m) \\
                 \sum_{m=2}^{\infty}m(m-1)\alpha_m\rho^{m-1}\cos(m\varphi +\delta_m) & V^{(1)\, 2}_2 \\
               \end{array}
             \right)
$$
with
$$
V^{(1)\, 1}_1\equiv V^{(0)}+\sum_{m=1}^{\infty}m\alpha_m\rho^{m-2}\sin(m\varphi +\delta_m)+\frac{\varepsilon}{2},
$$
and
$$
V^{(1)\, 2}_2\equiv V^{(0)}-\sum_{m=2}^{\infty}m(m-1)\alpha_m\rho^{m-2}\sin(m\varphi +\delta_m)+\frac{\varepsilon}{2}.
$$

Analogously, in three-dimensional space one can consider potentials with condition,
\be
\delta V=V^{(3)}-V^{(0)}=\triangle\chi =0.  \label{15}
\ee
Its solutions are expressed in terms of spherical harmonics:
$$
\chi =\sum_{j=0}^{\infty}A_jr^jY_j(\theta , \varphi),
$$
$$
Y_j(\theta , \varphi )=\sum_{m=0}^{j}\alpha_mP^m_j(\cos \theta )\sin (m\varphi +\delta_m),\,\,j=0,1,2, ...
$$
These functions allow to reconstruct the set of potentials $V^{(0)},\, V^{(1)\, k}_i,\, V^{(2)\, k}_i,\, V^{(3)}$ with specific properties of spectrum degeneracy. Usually, in the framework of Factorization Method $4\times 4$ Hamiltonians $H_B=\left(
                                                     \begin{array}{cc}
                                                       H^{(0)} & 0 \\
                                                       0 & H^{(2)} \\
                                                     \end{array}
                                                   \right)
$
and $H_F=\left(
                                                     \begin{array}{cc}
                                                       H^{(1)} & 0 \\
                                                       0 & H^{(3)} \\
                                                     \end{array}
                                                   \right)
$
are equivalent: their spectra coincide (supersymmetric degeneracy \cite{7}, \cite{8}). The condition (\ref{15})
leads to twofold degeneracy of a part of levels in $H_B$ and therefore, in $H_F,$ i.e. to fourfold degeneracy of a part of levels in the joint Hamiltonian (\ref{11}). Generally speaking, this additional degeneracy appears in three-dimensional case for a part of spectrum only, since some additional energy levels (common for $H^{(1)\, k}_i$ and $H^{(2)\, k}_i$) exist, which have no direct relation to the condition $\delta V=0.$\\

\noindent {\it Note added in translation.} We are grateful to our colleagues and, especially, to Juan Mateos Guilarte, for their interest to this old paper and for encouragement to translate it into English.

\end{document}